\documentclass{nature2007}
\usepackage{lineno}
\usepackage{fixltx2e}
\usepackage{amsmath}
\usepackage{textcomp}
\usepackage{graphicx}
\usepackage{cite}
\usepackage[a4paper]{geometry}
\usepackage{pdfpages}
\usepackage{placeins}

\title{Boundaries of the Peruvian Oxygen Minimum Zone shaped by coherent mesoscale dynamics}

\author{Jo\~ao H Bettencourt$^{1,2}$,
Crist\'obal L\'opez$^{1}$,
Emilio Hern\'andez-Garc\'\i a$^{1}$,
Ivonne Montes$^{3,6}$,
Jo\"el Sudre$^{4}$,
Boris Dewitte$^{4}$,
Aur\'elien Paulmier$^{4,5}$,
V\'eronique Gar\c{c}on$^{4}$
}

\makeatletter
\let\inserttitle\@title
\let\insertauthor\@author
\def\@cite#1#2{$^{\mbox{\scriptsize #1\if@tempswa , #2\fi}}$}
\makeatother

\begin{document}

\makeatletter
  \newpage\setlength{\parskip}{12pt}%
    {\Large\bfseries\noindent\sloppy \textsf{\@title} \par}%
    {\noindent\sloppy \@author}%
\makeatother




\begin{affiliations}
\item IFISC, Instituto de F\'isica Interdisciplinar
y Sistemas Complejos (CSIC-UIB), Campus Universitat de les Illes Balears,
E-07122 Palma de Mallorca, Spain. 
\item School of Mathematical Sciences,
University College Dublin, Dublin 4, Ireland.
\item GEOMAR, Helmholtz-Zentrum f\"ur Ozeanforschung Kiel,
Wischhofstr. 1-3, 24148 Kiel, Germany.
\item  LEGOS, Laboratoire d'Etudes en G\'eophysique et
 Oc\'eanographie Spatiales, 18, av. Edouard Belin, 31401 Toulouse Cedex  9,
France.
\item IMARPE, Instituto del Mar de Per\'u,
Esquina Gamarra y General Valle S/N, Chucuito, Callao, Per\'u.
\item IGP, Instituto Geof\'isico del Per\'u,
Lima, Per\'u.
\end{affiliations}

\section*{}
\textbf{Dissolved oxygen in sea water is a major factor affecting
marine habitats and biogeochemical
cycles\cite{Gnanadesikan2012, Lam2009,Ward2009}. Oceanic zones
with oxygen deficits represent significant portions of
the area and volume of the oceans\cite{Paulmier2009} and
are thought to be expanding\cite{Bopp2002,Stramma2008}. The
Peruvian oxygen minimum zone is one of the most pronounced and
lies in a region of strong mesoscale activity in the form of
vortices and frontal regions, whose effect in the dynamics of
the oxygen minimum zone is largely unknown. 
Here, we study this
issue from a modeling approach and a Lagrangian point of view,
using a coupled physical--biogeochemical 
simulation of the Peruvian oxygen minimum zone and finite-size
Lyapunov exponent fields to understand the link between mesoscale dynamics
and oxygen variations.
Our results show that, at depths between
380 and 600 meters, mesoscale structures have a relevant dual
role. First, their mean positions and paths delimit and
maintain the oxygen minimum zone boundaries. Second, their high
frequency fluctuations entrain oxygen across these boundaries
as eddy fluxes that point towards the interior of the 
oxygen minimum zone and are one order of magnitude 
larger than mean fluxes. We conclude that these eddy fluxes 
contribute to the ventilation of the oxygen minimum zone.}

Regions of the ocean with strong O$_2$ deficiency in the water
column are called Oxygen Minimum Zones (OMZs). The OMZs are
differentiated in the vertical by three distinct layers: the
oxycline (upper O$_2$ gradient), the core (typically with $O_2
< 20 \ \mu M$) and the lower O$_2$ gradient.
The Eastern Tropical South Pacific (ETSP) contains one of the
three major permanent OMZ, with an oxycline extending
from the upper $10$ to $170$ meters, 
a core with a
thickness of around $400$ meters\cite{Paulmier2009} and lower
oxygen gradient 
extending to
about $3700$ meters\cite{Paulmier2009}. This OMZ is maintained
by the combination of significant rates of biological
production and decomposition of sinking organic
material\cite{Chavez2009} at the Peruvian upwelling region, and
weak circulation in the shadow zone of the 
southern
Pacific subtropical gyre. The circulation is then dominated by
the equatorial and eastern boundary current
systems\cite{Penven2005,Montes2010}. Energetic vortices, called
mesoscale eddies, and filaments are ubiquitous in this
area\cite{Chaigneau2008}. In contrast to the oligotrophic
regions of the ocean where mesoscale eddies can sustain
biological productivity\cite{Oschlies1998}, in upwelling
regions stirring by eddies -- the process of tracer gradient
intensification by advection -- tends to inhibit biological
production\cite{Rossi2009,Gruber2011}.

In the ETSP, the role played by mesoscale structures in the
distribution of O$_2$ within the OMZ remains unclear and we
approach this issue by analyzing data from a coupled
physical-biogeochemical high-resolution model\cite{Montes2014}
of the regional ETSP (see Methods), and characterizing
mesoscale transport and stirring by means of Finite-size
Lyapunov exponent (FSLE) fields\cite{Aurell1997,dOvidio2004}
(see Methods and Supplementary Information). Maxima in these
fields form thin filamentary structures, the so-called
Lagrangian Coherent Structures
(LCS)\cite{Haller2000b,dOvidio2004,dOvidio2009} identifying the
most intense mesoscale regions and acting as barriers for fluid
transport across them.

In this work we focus on the transport aspects of the
mesoscale-OMZ interaction, particularly in the OMZ boundaries,
and the fluxes across them. We do not address specifically the
biogeochemical processes occurring inside the zone which are
certainly determinant (and are included in our
regional simulation model) but we gauge instead the physical
effects of the mesoscale structures on the OMZ dynamics. This
is done by: a) computing correlations between the (temporally
averaged) O$_2$ concentration and FSLE at layers located at
different depths; b) studying events of O$_2$ rich-waters
entrainment into the OMZ; c) calculating the temporal average
of O$_2$ normal fluxes across the northern and southern
boundaries of the OMZ as a function of depth, and its
correlation with the average mixing measurement obtained from
FSLE. 
The outcome of these analyses is that, despite the important 
biogeochemical processes, mesoscale stirring already 
determines many important features of the oxygen distribution.

\begin{figure}
\centering
\includegraphics[width=\textwidth]{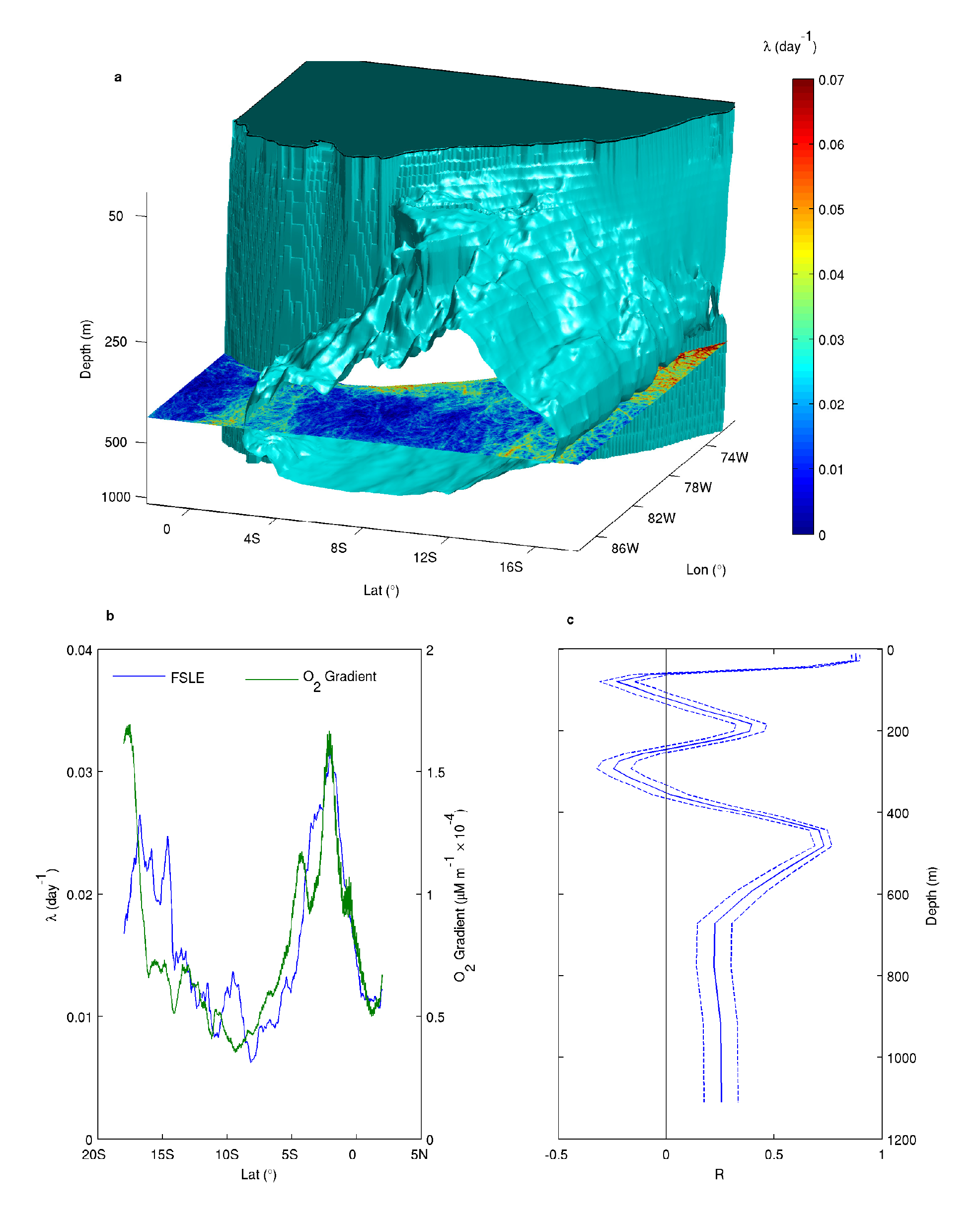}
\caption{OMZ core, finite-size Lyapunov exponent (FSLE) and FSLE-O$_2$ gradient 
correlations for simulation year 21. a) 20 $\mu M$ isosurface of mean O$_2$ concentration 
and the mean FSLE field at 410 meters depth. Note the stretched vertical scale. 
Flat top shows the Peruvian coast. b) Zonally averaged mean (z.a.m.)
FSLE and O$_2$ gradient, averaged between 380 m and 600 m depth. c) Pearson
correlation coefficient ($R$) between z.a.m. FSLE and O$_2$ 
gradient (solid line). Fisher 95\% confidence interval on R (dashed line).}
\label{fig:fsle3d}
\end{figure}

The $20 \ \mu M$ isosurface of the annual mean O$_2$ field for
simulation year (s.y.) 21 (Fig. \ref{fig:fsle3d}a) gives an
OMZ core with maximal horizontal extension at approximately
$400$ m depth extending between $3^o$ S and $16^o$ S. The
higher O$_2$ concentrations north of $2^o$ S are associated
with eastward equatorial subsurface currents carrying
relatively oxygen-rich water\cite{Stramma2010,Czeschel2011},
while the southern increase of O$_2$ ($14^o$ S to $17^o$ S) is
adjacent to the northern part of the subtropical gyre. Figure
\ref{fig:fsle3d}a also displays the annual mean backward FSLE
field at $410$ m depth, which shows a high correlation with the
mean O$_2$ field delineating the limits of the OMZ core. The
FSLE mean field is structured as zonal bands coincident with
the north and south OMZ boundaries with relatively high FSLE
values when compared to the core region. Both bands signal
stirring by the eddies formed 
at the continental shelf and advected offshore, and by other
mesoscale processes\cite{Czeschel2011,Chaigneau2008}. This
indicates that the enhanced mesoscale activity in those areas
delineates the limits of the average OMZ core region.

Since LCS (that we locate as maximum values of FSLE) act 
as transport barriers,
large gradients of O$_2$ should occur across them 
\cite{Lehahn2007}. Thus we expect to find a relationship
between the stirring intensity as measured by the FSLE and the
O$_2$ gradient norm (see additional discussion in Supplementary
Information; in the following, the term O$_2$ gradient
refers to the norm). The
relationship between both fields is quantified in Fig.
\ref{fig:fsle3d}b where we plot the latitudinal variation between 18$^o$ S and 2$^o$ N
of mean FSLE and O$_2$ gradient averaged 
between the coast and 85$^o$W and from 380 to 600 m depth,
showing the coincidence in the maxima of both quantities: the
maxima of FSLE indicating the positions of the LCS and the
maxima of the O$_2$ gradient signalling the northern and
southern boundaries of the OMZ. This correlation is not equally
strong at all depths as it is shown in Fig. \ref{fig:fsle3d}c,
where we plot the vertical profile of the Pearson correlation
coefficient \cite{sheskin-2003-handbook} , $R$, between the zonally averaged
mean FSLE and mean O$_2$ gradient. Roughly, we can distinguish
two areas in the OMZ core: a) between $190-350$ meters where
these quantities show correlations of alternating sign; and b)
between $380-600$ meters where the correlation is large and
positive (with an average $R$ of $0.748$).
It is in this subregion of the OMZ core that its boundaries 
are strongly determined by the mesoscale
dynamics. This is so because the OMZ dynamics is a balance between
hydrodynamic and biogeochemical processes. At these depths,
the physical forcing, albeit lower than in the upper layers,
has a variability two orders of magnitude larger than the biogeochemical forcing
\cite{Montes2014} (see Supplementary Information) and its effects 
are clearly visible in the
strong correlation between FSLE and O$_2$ gradients.

\begin{figure}
\centering
\includegraphics[width=\textwidth]{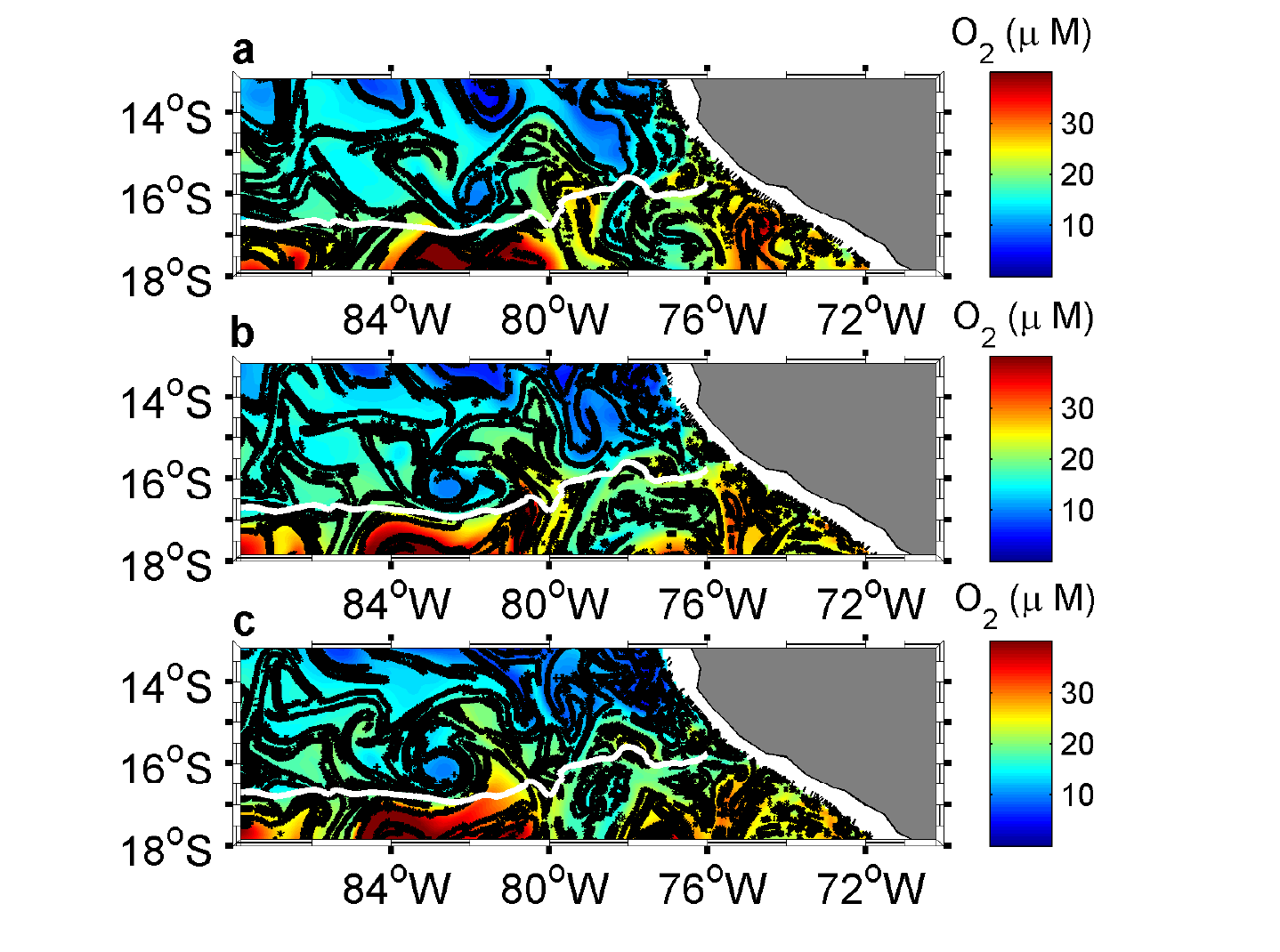}
\caption{Entrainment of O$_2$ rich waters into the OMZ due to the motion of
Lagrangian coherent structures.
Color codifies O$_2$ at $410$ m depth and the lines are the $0.075 \ day^{-1}$
FSLE isolines. a) $16$ September;
b) $7$ October; c) $25$ October; all of s.y. 21. Note the oxygen-rich tongue entering
the OMZ at $80-82^{o}$W. White continuous line is
the $20\ \mu M$ mean isoline at $410$ m depth (corresponding to the southern OMZ boundary).
}
\label{fig:eddypassage}
\end{figure}

Besides mean behaviour, individual events are also relevant,
since mesoscale eddies are able to transport waters with
different biogeochemical properties with respect to surrounding
areas, 
giving rise to sporadic episodes of high O$_2$
patches inside the OMZ. Fig. \ref{fig:eddypassage} shows
one of these temporal sequences where an eddy dipole (with
borders signalled by maxima of FSLE at $410$ meters) entrains
water with high oxygen content (the red-yellow tongue at
$80-82^{o}$W) towards the interior of the OMZ. This episode had
a duration of approximately $3$ months 
(9/9 to 1/12 of s.y. 21; first month displayed)
and during this period the entrainment of these waters
carried $0.4\times10^6$ mol of O$_2$ per meter of depth into
the OMZ at this depth. Episodes of this nature are frequent at
the southern boundary although often less intense (see
Supplementary Information). At the northern boundary the
frequency of O$_2$ injection episodes is higher and they last
longer.
The difference between both boundaries rests upon the spatial
distribution of O$_2$. Since the mean northern boundary is
almost coincident with the large O$_2$ gradient region most of
the time, any small displacement of this region will cause a
significant change in the O$_2$ signal right at the boundary.
In the southern border gradients are more distributed and
strong anomalous O$_2$ signals will be caused only by the mesoscale
eddies entraining water across the boundary as in Fig.
\ref{fig:eddypassage}. 
At both boundaries the episodic ventilation of the OMZ follows
an offshore path consistent with the propagation of eddies and
other perturbations from the coastal waters to the open ocean.
A 
characterization of FSLE and O$_2$
joint dynamics in terms of wavelet spectra,
emphasizing the dominant periods, is 
presented in the
Supplementary Information.

\begin{figure}
\centering
\includegraphics[width=.9\textwidth]{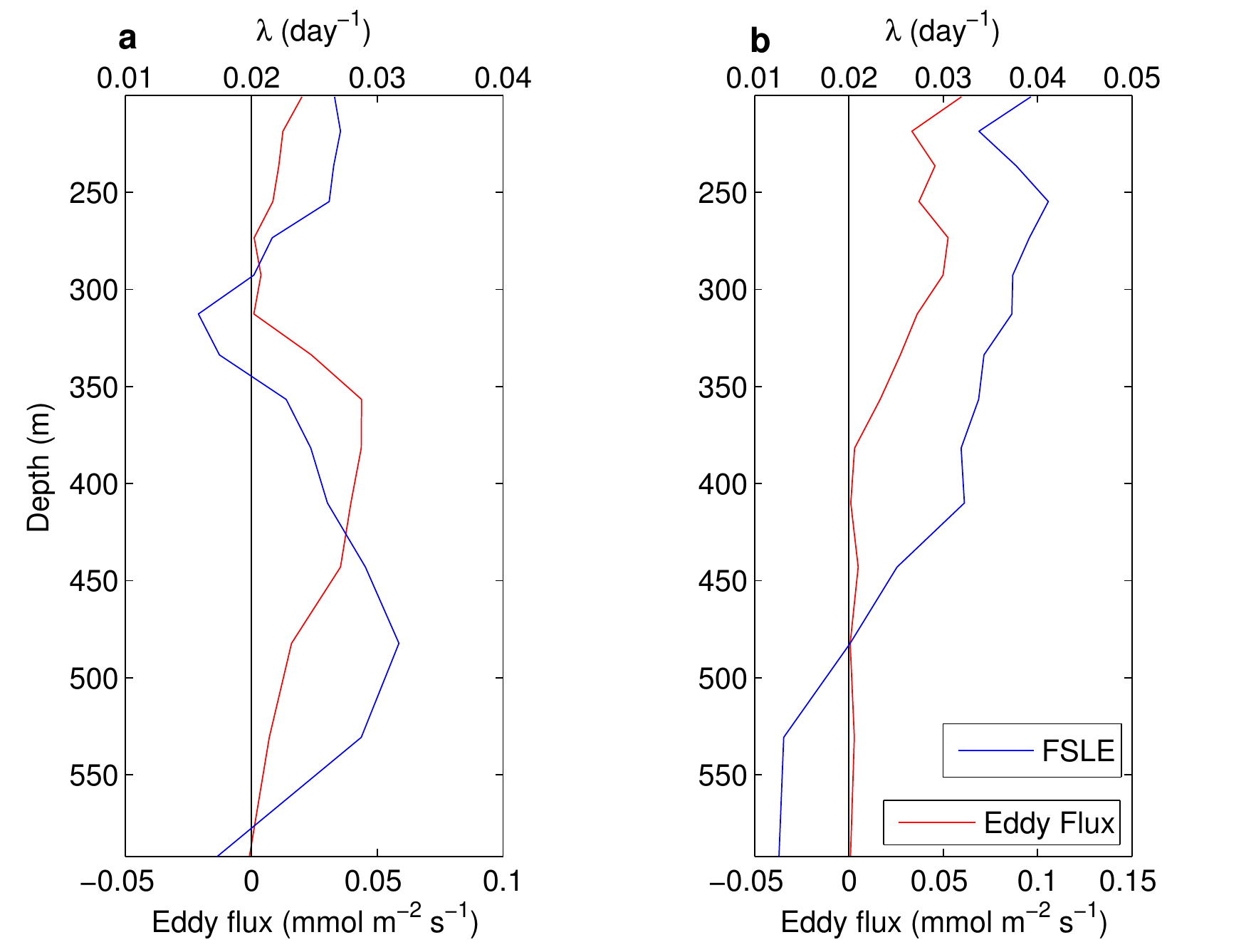}
\caption{Vertical profiles of FSLE and O$_2$ eddy fluxes from 200 to 600m. Mean FSLE (blue lines) and normal
O$_2$ eddy flux (red lines) averaged from the coast until $88^{o}$ W for each depth and during
simulation year 21 at (a) Northern $20\ \mu M$ mean boundary and (b) Southern $20\ \mu M$ mean boundary.
Positive fluxes bring O$_2$ into the OMZ. 
The vertical line indicates the zero value for the fluxes.
}
\label{fig:eddyfluxesFSLE}
\end{figure}

The average amount of
O$_2$ entering through the OMZ boundaries due to mesoscale
processes was quantified
by 
computing eddy fluxes of
O$_2$ normal to the northern and southern limits. 
Small-scale turbulent diffusion produces much smaller fluxes. 
Eddy fluxes were calculated across the mean $20 \ \mu M$ level
boundary between $200$ and $600$ meters of depth during s.y.
21, from the covariance between velocity
anomalies and O$_2$ concentration anomalies (see Supplementary
Information). Vertical fluxes across these borders were always
orders of magnitude smaller than horizontal ones (see
Supplementary Information), and thus the following discussion
about horizontal components does also apply to the total normal
flux. 
At the northern boundary the horizontal eddy flux
profile is mainly positive (Fig. \ref{fig:eddyfluxesFSLE}a, red
line), meaning that the O$_2$ variance due to horizontal eddy
fluxes is bringing O$_2$ into the OMZ. The highest eddy fluxes
are reached at core depths between $350$ and $500$ meters which
is close to the depth range where the higher FSLE mean values
at the boundary are obtained (Fig. \ref{fig:eddyfluxesFSLE}a,
blue line), although the maximum value of this latter quantity
(associated to the presence of the subsurface
currents\cite{Stramma2010,Czeschel2011}) appears deeper than
the eddy flux maximum ($350$ vs $480$ meters). Above $300$
meters the horizontal eddy fluxes are small and the minimum is
obtained around $300$ meters, where the FSLE is also minimum.
Globally integrated between $200$-$600$ m depth and along the
northern $20\ \mu M$ boundary from coast until $88^{o}$W, the
horizontal eddy flow rate towards the OMZ interior is of
$6.15\times10^6\mu mol~ s^{-1}$, whereas the corresponding mean
flow rate is $2\times10^5\mu mol~ s^{-1}$ and directed outwards
(see Supplementary information). At the southern $20 \ \mu M$
mean boundary, eddy fluxes are also positive (
Fig. \ref{fig:eddyfluxesFSLE}b) along the range of depths
considered, being fairly constant from $200$ to $300$ m, and
nearly vanishing between $400$ to $600$ m depth.
Integrating between $200$-$600$ m  from coast to $88^{o}$W, the
eddy flow rate towards the OMZ interior is of $1\times10^7\mu
mol~ s^{-1}$, whereas the mean flow rate is $3.56\times10^5\mu
mol~ s^{-1}$ and directed outwards (see Supplementary
Information).

The differences in the O$_2$ eddy fluxes between the northern
and southern boundaries may be understood in terms of the
mesoscale activity. In the southern boundary O$_2$ anomalies
are caused by eddies (signalled by large FSLE) crossing the
boundary. Thus higher O$_2$ eddy fluxes should be associated
with higher FSLE values, which indeed is true looking at the
profiles in Fig.\ref{fig:eddyfluxesFSLE}b). On the northern
boundary,  this holds until the local minimum at 312 m. Below
this depth, O$_2$ anomalies crossing the northern border are
mainly related to, as stated above, fluctuations in the
position of a large O$_2$ gradient zone, associated to
fluctuations in the subsurface equatorial currents, separating
the subsurface O$_2$ rich equatorial waters from the OMZ core.

To conclude, in this work we have addressed the role of
mesoscale structures that populate the OMZ in the ETSP.
We identified the boundaries of these mesoscale eddies and fronts
as LCS that act as barriers to transport controlling fluid
interchange in and out the OMZ. Comparison of the FSLE approach
with an exit time characterization (see Supplementary
Information) supports this view. Despite the important
biogeochemical processes, mesoscale stirring already shapes
important features of the oxygen distribution. We find that
mesoscale dynamics plays a dual role, which can be respectively
associated with the average behaviour and with the turbulent
fluctuations. The northern and southern boundaries of the OMZ
core are well determined by the averaged mesoscale dynamics, in
particular for depths between $300$ and $600$ meters, where a
good correlation between mean FSLE and O$_2$ gradients was
found. At other depths the relation between FSLE and O$_2$ may
not hold, indicating significant O$_2$ forcing by biogeochemical
processes. Episodic events of OMZ ventilation are produced by
eddy stirring where waters with high O$_2$ content
are entrained into the OMZ by the action of mesoscale eddies.
On the whole, between $200$ and $600$ m depth, eddy fluxes were
found to bring O$_2$ inside the OMZ at both the northern and
southern frontiers, while O$_2$ mean fluxes were much smaller
and in the opposite direction. The biogeochemical processes
occurring in the interior of the OMZ would provide the dominant
oxygen consumption sink to close the O$_2$ budget and maintain
the OMZ core.


\begin{methods}

\section*{Circulation and OMZ modeling} \label{sec:met-circ}

The circulation and OMZ modeling in the Eastern Tropical
Pacific was accomplished by the combination of the hydrodynamic
model ROMS\cite{Shchepetkin2005} (Regional Ocean Modeling
System) and the biogeochemical model
developed\cite{Gutknecht2013} for the Eastern Boundary
Upwelling Systems (BioEBUS). The Eastern Tropical Pacific
configuration covers the region from $4^o$ N to $20^o$ S and
from $70^o$ to $90^o$ W with an horizontal resolution of
$1/9^o$ degrees ($\approx 12$ km) and $32$ terrain-following
vertical levels with variable vertical resolution (higher in
the upper ocean). The coupled model is run in a climatological
configuration previously validated\cite{Montes2011} for the
Eastern Tropical South Pacific, and the present configuration
has been recently validated and a sensitivity analysis
performed\cite{Montes2014}. The model was forced by the
QuickSCAT\cite{Liu1998} wind stress monthly climatology and by
heat and fresh water fluxes from the
COADS\cite{daSilva1994atlas} monthly climatology. The dynamical
variables at the three open ocean boundaries are provided by a
monthly climatology computed from the Simple Ocean Data
Assimilation reanalysis\cite{Carton2008}. For the
biogeochemical model, boundary conditions of nitrate and oxygen
concentrations are taken from CSIRO Atlas of Regional Seas
(CARS 2009, http://www.cmar.csiro.au/cars) and chlorophyll a
concentration from SeaWiFS (http://oceancolor.gsfc.nasa.gov/).
The simulations were performed for a 22-year period. The first
$13$ years were run with the physics only and the following
9-years were run with the physical/biological coupling. The
coupled model reached a statistical equilibrium after $4$ years
and model outputs were then stored every $3$ days (averaged).

\section*{Finite-size Lyapunov exponent} \label{sec:met-fsle}

The Finite-size Lyapunov exponent (FSLE), $\lambda$, is a
measure of the rate of divergence in the positions of particle
pairs while separating from an initial distance $\delta_0$ up
to a final distance $\delta_f$. It was developed to study
non-asymptotic dispersion processes\cite{Aurell1997} and to
quantify dispersive behavior of particles, especially in those
cases where length scales are easier to identify than temporal
ones. It is given by the following expression:
\begin{equation*}
\lambda =\frac{1}{\tau}\log{\frac{\delta_f}{\delta_0}},
\label{eq:fsle}
\end{equation*}
where $\tau$ is the time needed for the initial separation to
increase from $\delta_0$ to $\delta_f$. The FSLE is a function
of the initial and final separations, and also of the initial
location of the particle pair ${\bf x_0}$ and of the time of
release $t_0$. Thus, the computation of $\lambda$ for a given
set of initial locations and in a time interval provides an
insight to the locations of weaker/stronger particle dispersion
and its evolution with time in the domain $D$. In fluid flows,
regions that exhibit substantial stretching of fluid material,
hence high values of $\lambda$, have filamental shapes and have
been associated with barriers and avenues to
transport\cite{dOvidio2004} that strongly constrain the mixing
of fluid with different properties, the so-called LCS
\cite{Haller2000b,dOvidio2004,dOvidio2009}. Trajectory
integration can be done from the present to the future, forward
in time, or towards the past, backwards in time. The locations
with high values of the backwards Lyapunov field are the
structures better delimiting the distribution of transported
substances and providing barriers to
transport\cite{dOvidio2004,dOvidio2009}. Thus, the FSLE
backwards field is the one used in this paper.

To compute the three-dimensional FSLE field we extended a
previous two-dimensional method\cite{dOvidio2004} to include
the third dimension, by computing the time $\tau$ it takes for
particles initially separated by $\delta_0=(\varDelta x_0^2 +
\varDelta y_0^2 + \varDelta z_0^2)^{1/2}$ to reach a final
distance of $\delta_f=(\varDelta x_f^2 + \varDelta y_f^2 +
\varDelta z_f^2)^{1/2}$. However, in a similar application for
the Benguela upwelling system\cite{Bettencourt2012} it was
observed that the displacement in the vertical $z$ direction
does not contribute significantly to the calculation of
$\delta_f$ and so we define a quasi-3d computation of FSLE: we
use the full three dimensional velocity field for particle
advection but particles are initialized in horizontal ocean
layers and the contribution $\varDelta z_f$ is not considered
when computing $\delta_f$.
\end{methods}

%
%
%

\section*{}

%

\begin{addendum}
 \item[Correspondence] Correspondence and requests for materials
 should be addressed to JHB.\\~(email:joao.bettencourt@ucd.ie).

 \item
 JHB, CL and EHG acknowledge support from FEDER and MINECO
 (Spain) through projects ESCOLA (CTM2012-39025-C02-01) and 
 INTENSE@COSYP (FIS2012-30634). JHB acknowledges financial support 
 of the Portuguese FCT (Foundation for Science and Technology) and Fundo
 Social Europeu (FSE/QREN/POPH) through the predoctoral grant 
 SFRH/BD/63840/2009. IM would like to acknowledge the EUR-OCEANS 
 Consortium for support through a Flagship post-doctoral fellowship 
 to IM on deoxygenation in the oceans. 
 \item[Competing Interests] The authors declare that they have no
 competing financial interests.

 \item[Author Contributions]
 JHB, CL, EHG, BD and VG directed the study; JHB, CL, EHG, BD, IM, JS, AP,
 VG analyzed data and performed numerical simulations;
 JHB, CL, EHG, VG, wrote the paper with significant contributions from BD.

\end{addendum}

\newpage
\newgeometry{left=1cm,right=1cm,bottom=0.1cm,top=0.1cm}
\restoregeometry

\newpage

\newpage

\FloatBarrier



\section*{Supplementary material}

# Boundaries of the Peruvian Oxygen Minimum Zone shaped by coherent mesoscale dynamics.

## Supplementary Information

by João H. Bettencourt[*,1,2], Cristóbal López[1], Emilio Hernández-García[1], Ivonne Montes[3,6], Joël Sudre[4], Boris Dewitte[4], Aurélien Paulmier[4,5], and Véronique Garçon[4]

[1] IFISC, Instituto de Física Interdisciplinar y Sistemas Complejos (CSIC-UIB),
Campus Universitat de les Illes Balears, E-07122 Palma de Mallorca, Spain.
[2] School of Mathematical Sciences,
University College Dublin, Dublin 4, Ireland.
[3] GEOMAR, Helmholtz-Zentrum für Ozeanforschung Kiel,
Wischhofstr. 1-3, 24148 Kiel, Germany.
[4] LEGOS, Laboratoire d'Etudes en Géophysique et Océanographie Spatiales,
18, av. Edouard Belin, 31401 Toulouse Cedex 9, France.
[5] IMARPE, Instituto del Mar de Perú,
Esquina Gamarra y General Valle S/N, Chucuito, Callao, Perú.
[6] IGP, Instituto Geofísico del Perú, Lima, Perú.

## Backward FSLE three-dimensional fields

Here we give additional details on the calculation of the finite-size Lyapunov exponent (FSLE) fields and further discuss their horizontal and vertical structure. Maximum values of FSLE computed backwards in time can be interpreted as fronts of advected tracers, since they delineate the lines along which the tracers are stretched and folded by the fluid flow. By the same reason, they provide barriers to transport with little or no flow across them [1, 2].

We computed daily three dimensional ($3d$) fields of backward FSLE for simulation year 21. This was done by setting up a regular $3d$ latitude-longitude-depth

grid of initial conditions with spacing in the two horizontal directions of $\delta_0 = 1/27^o$ (i.e. $\approx 4$ km) and 30 vertical layers with variable vertical spacing. The grid covered the eastern tropical south Pacific (ETSP) from $88^o$ W to $70^o$ W and $18^o$ S to $2^o$ N. Vertically, the grid extended from 10 to 1100 m depth and was clustered in the upper 500 m of ocean. To obtain the FSLE field, $\lambda(\mathbf{x}, t)$, at location $\mathbf{x}$ and time $t$, one particle is released at time $t$ from grid node at location $\mathbf{x}$ and the separation to the particles released from neighboring nodes is monitored. $\tau$ is the time needed for the first of these separations to reach the value $\delta_f$ (that in this study was set at 100 km). Trajectories were integrated backward in time for 6 months with a Runge-Kutta $4^{th}$ order method. If at the end of this interval the separations are smaller than $\delta_f$, or if the particles leave the domain or hit the shore, then the FSLE for the release location is set to zero. Otherwise, the FSLE is computed as

$$\lambda(\mathbf{x}, t) = \frac{1}{\tau} \log \frac{\delta_f}{\delta_0} \ .$$

A map of instantaneous backward FSLE is shown in Fig. S1 for different depths. In the upper layer (Fig. S1 a) we observe that south of $4^o$ S the FSLE field is organized in thin filamental features with high FSLE values superimposed on a low FSLE background. North of $4^o$ S, we see a larger density of small-scale features, that represent the change in dynamical regime [3] as we approach the equator. Time scales associated with the Lagrangian dynamics at this depth vary from 4 to 10 days (the FSLE is roughly an inverse time scale). At larger depths (Fig. S1 b, c and d) the features of the instantaneous FSLE field are maintained although the intensity is reduced from its values at the surface. The area near the equator with dense small-scale structure is reduced at 113 m depth and qualitatively different and shifted south at 410 and 592 m depth (Fig. S1 b, c and d, respectively). At all depths

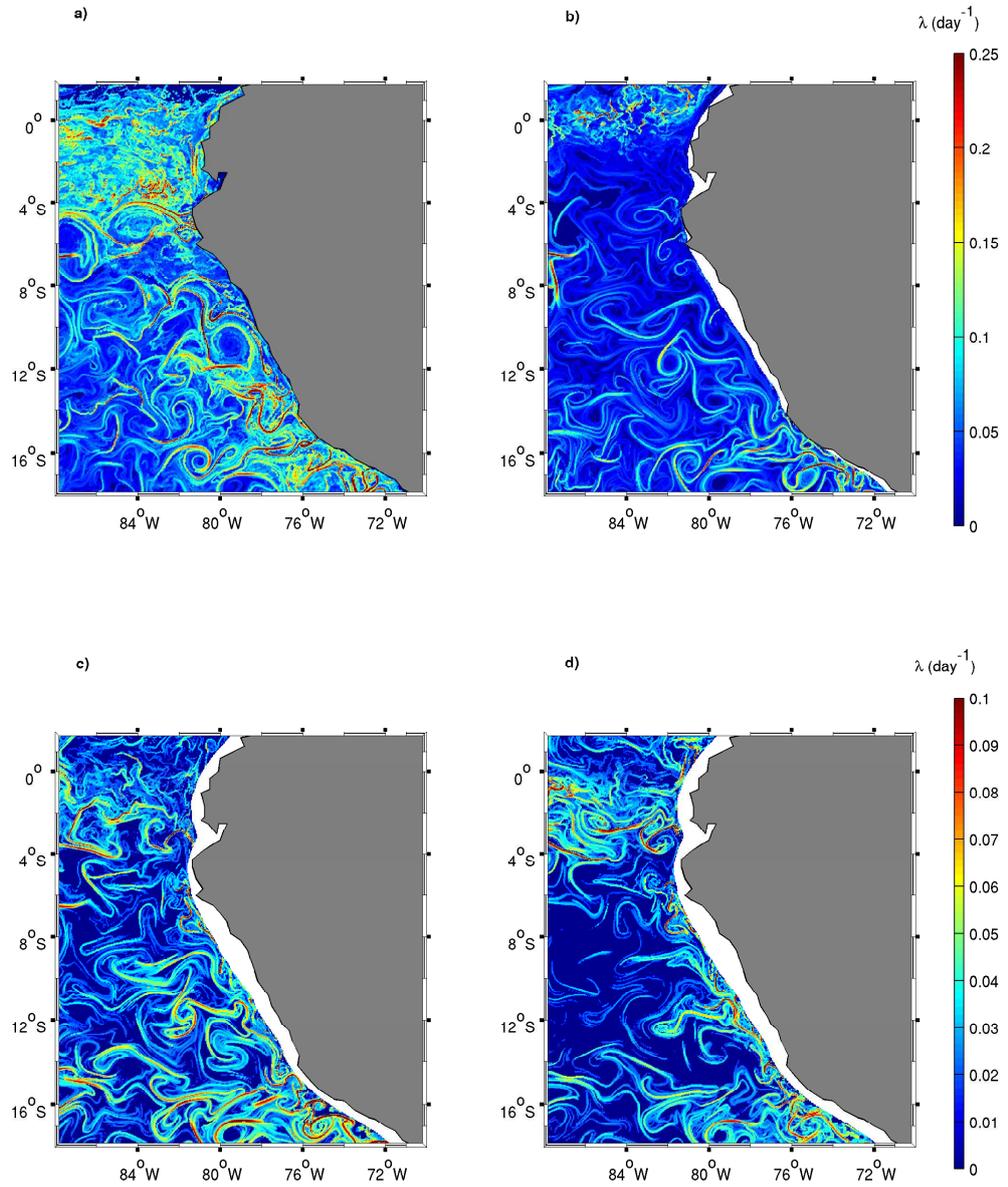

Figure S1: Map of backward FSLE for the $1^{st}$ of January of simulation year 21. a) 10 m depth. b) 113 m depth. C) 410 m depth. D) 592 m depth. Note the different colorbar at different depths.

3

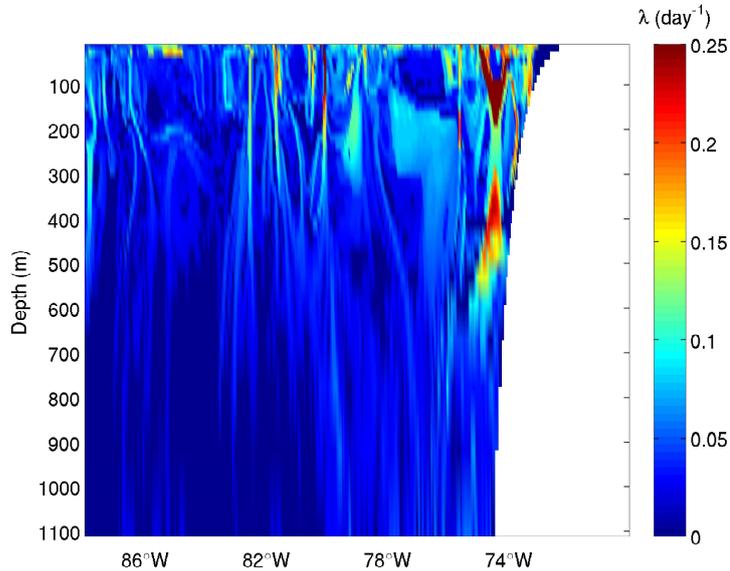

Figure S2: Vertical map of backward FSLE for the $1^{st}$ of January of simulation year 21, at $16.45^o$ S.

shown, the high FSLE lines are more common near the coast, which in this region is the main source of mesoscale variability due to instability of coastal currents and the associated upwelling regime [4, 5].

The vertical structure of the instantaneous FSLE field is shown in Fig. S2 for the same date on a vertical section at $16.45^o$ S. The thin filamental structures appearing in the horizontal maps have a vertical extension of about 400 m from the surface down. The structures oriented parallel to the zonal axis reveal this *curtain-like* shape, as can be observed between $76^o$ W and $78^o$ W and 100 and 300 m depth. Such shape has already been observed (in $3d$) in similar calculations for the Benguela upwelling zone [6] and can be understood theoretically [7]. Dynamical studies with particle trajectories showed that the most intense *curtains* can be related to Lagrangian eddy boundaries.

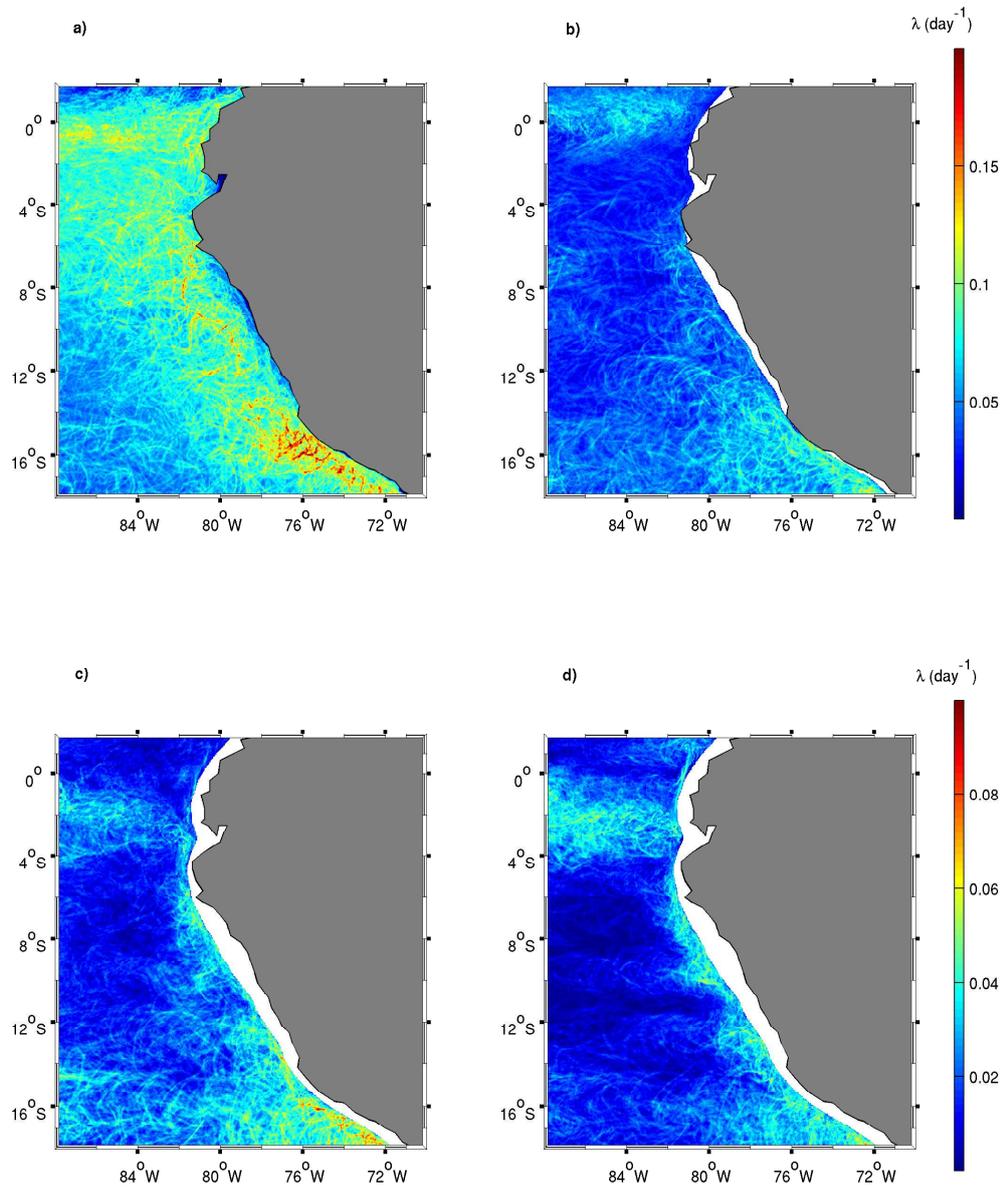

Figure S3: Horizontal maps of backward FSLE temporally averaged for simulation year 21. a) 10 m depth. b) 113 m depth. C) 410 m depth. D) 592 m depth. Note the different colorbar at different depths.

5

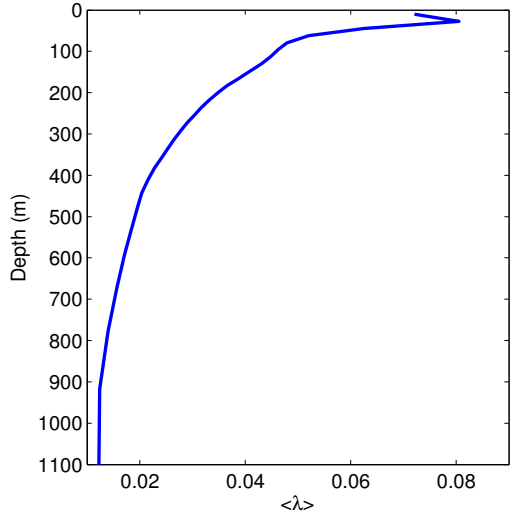

Figure S4: Time mean of the horizontally averaged depth profile of the backward FSLE field $<\lambda>$ (in day$^{-1}$) for year 21.

The temporal-mean horizontal maps of the backward FSLE fields are shown in Fig. S3 for four depths: 10, 113 and 410 and 592 m. One can observe the most persistent patterns of mixing, in particular, the northern and southern strips delineating the boundaries of the oxygen minimum zone (OMZ). The depth profile of horizontally averaged mean FSLE field for simulation year 21 (Fig. S4) shows the decrease in stirring with depth, with a subsurface peak at $\approx 30$ m depth. This decrease of stirring with depth, as measured by the FSLE, was also found in a study of the Benguela upwelling region [6]. Its origin could be simply the overall smaller velocities found at deeper layers, and also the decrease in the nonlinearity of the mesoscale eddies (as indicated for example by its ratio between rotation and propagation velocities) that occurs with depth [5].

# Residence times

Residence times (RT) are a tool to study fluid exchange between two regions. Although the OMZ does not constitute a geographically determined region such as a basin or bay, it is still possible to define appropriate limits to the OMZ (as we have done all along this paper) so that fluid exchange between the OMZ and its exterior may be studied with RT distributions [8, 9]. They are a Lagrangian diagnosis complementary to the FSLE methods. The RT of a fluid particle is defined as the time it spends inside a certain region before crossing a particular boundary. Here we compute the RT in the OMZ as the time the particle remains with a $O_2$ content below 20 $\mu M$. The results obtained are similar if instead we compute the exit time from the average OMZ core region, i.e. the volume in which the temporally averaged $O_2$ concentration is smaller than 20 $\mu M$. Trajectories can be integrated forward in time (and then the computed time is properly an exit time) or backward in time (so that one is calculating then the time the particle has been in the region before the present moment).

We show in Fig. S5 the RT distribution at 410 meters depth for the $1^{st}$ of July of simulation year 20. Similar results are observed at other depths. To obtain this distribution, particles were released from the same horizontal regular grid described in the previous section and the trajectories were integrated backward and forward for 2.5 years (the need to use such large integration times forced us to use as initial time the simulation year 20 instead of year 21 as in other calculations, since the simulation dataset contains only 22 years). Backward and forward FSLE fields were also computed for this date and location for comparison purposes.

The RT are larger than 2.5 years in most of the area identified as the OMZ core. In contrast with this rather homogeneous distribution in the centre, at the

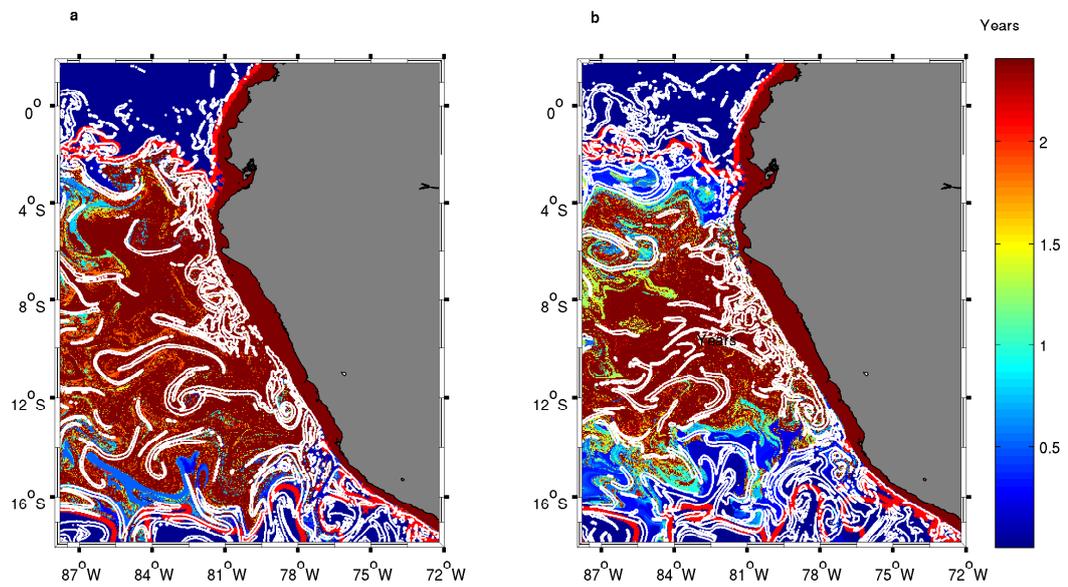

Figure S 5: Residence times (RT) at 410 m depth for $1^{st}$ July of simulation year 20. a) Backward RT. Superimposed white lines are 0.035 $day^{-1}$ backward FSLE isolines. b) Forward RT. Superimposed white lines are 0.038 $day^{-1}$ forward FSLE isolines. The magenta lines bound the region inside which the concentration of $O_2$ was smaller than $20\mu M$ already at the initial time, so that outside these lines the exit times are zero. The red color associated to a residence time of 2.5 years is in fact associated also to all residence times larger than this duration.

OMZ boundaries the RT distribution exhibits quite complex shapes, with several re-entrant zones of low RT and sharp transitions to low RT. Additionally, there are several cases of thin regions of low RT intruding in to the high RT central zone. These sharp boundaries between high and low RT coincide in some areas with high FSLE values. This coincidence is a feature observed with the RT distributions of passive particles in a given geographical region [9]. Since passive particles are Lagrangian tracers, we conclude that in the areas of matching RT changes and FSLE lines, the $O_2$ content is conserved along particle trajectories and local changes in $O_2$ occur mostly due to advection. The long residence times found are consistent with the low ventilation regime accompanied with biogeochemical processes which reduce oxygen concentrations to hypoxic levels.

## Horizontal $O_2$ gradients

Here we enlarge our discussion on the relationship between FSLE and $O_2$ horizontal gradients. First we observe in Fig. S6 a) how the lines of high FSLE values determine instantaneous (at depth 410 m) fronts for the $O_2$ concentrations. These are located in the northern and southern boundaries of the OMZ as has been discussed all along the text. In Fig. S6 B) it is shown how the high-FSLE lines coincide with the largest horizontal gradients of $O_2$.

In Fig. S7 a) and b) we show FSLE and $O_2$ gradient maps, respectively, averaged temporally (for simulation year 21) and vertically in the mid-depth range of the OMZ (between 350 and 600 m). We do not discuss the relationship between the high values of FSLE and of $O_2$ gradients found very close to the coast, since that region has very distinct dynamics and biogeochemistry, and it is influenced by the presence of

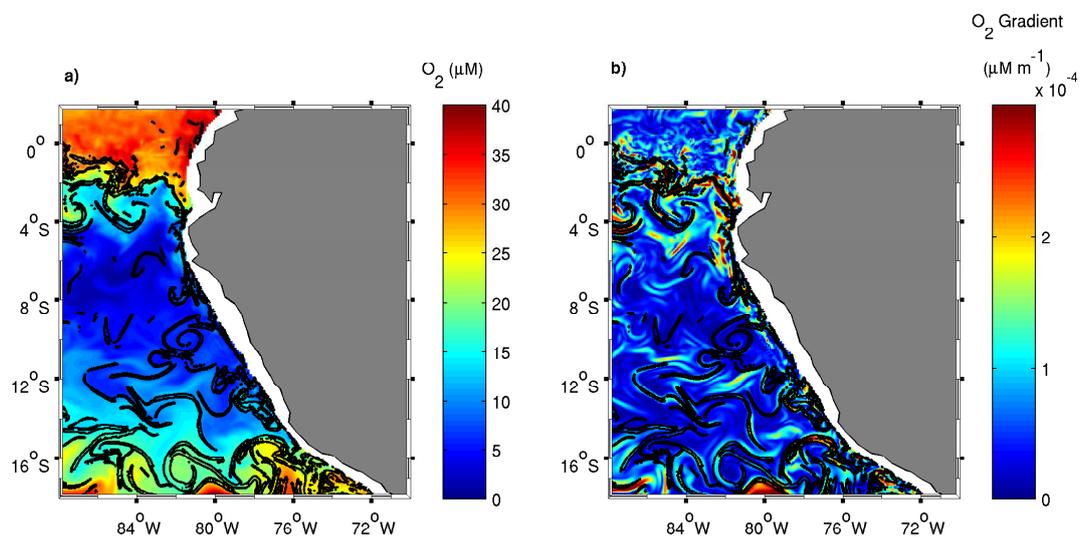

Figure S 6: Lines of high FSLE values ($> 0.04\ day^{-1}$, black) superimposed on instantaneous $O_2$ and $O_2$ gradient fields on $1^{st}$ of March of year 21 at 410 m depth. a) $O_2$ concentration. b) $O_2$ horizontal gradient.

coastal currents and water-column sediments interaction. The maps clearly show the high gradient zones north and south of a central basin with low average $O_2$ gradient that coincides with the OMZ. The northern band is located approximately at an interface between an eastward mean zonal flow relatively rich in $O_2$ [10, 11] and a southern adjacent westward mean flow. This flow configuration at middepth present in our simulation data shows similarities with ADCP data [12] taken in February 2009 that shows the SICC (South Intermediate Counter Current) eastward flow and the adjacent SEIC (South Equatorial Intermediate Current) westward flow with velocities about 2.5 $cm/s$. The mean FSLE field for the same depth range (Fig. S7 a) shows that the band between $0^o$ and $4^o$ S is a zone of high FSLE. On the southern boundary, high $O_2$ gradients are located close to the equatorward edge of the southern subtropical gyre, along a zonal band below $16^o$ S. In this region, we also find high values of the mean FSLE but these are distributed along a wider zonal region, poleward from $14^o$ S.

The center of the OMZ forms a basin of low $O_2$ gradients and also low FSLE. The low values of $O_2$ are due to low or inexistent ventilation (as quantified by large residence times) of this region coupled to the respiration of sinking organic matter. The low FSLE distribution is due to the absence of significant mesoscale activity at this depth. Although this region is located offshore of the Peru upwelling strip, and receives anticyclonic eddies formed by instability of the coastal currents [13], mesoscale activity as measured by eddy intensity decreases as we move offshore [5], and it is only intense in the two strips surrounding the OMZ.

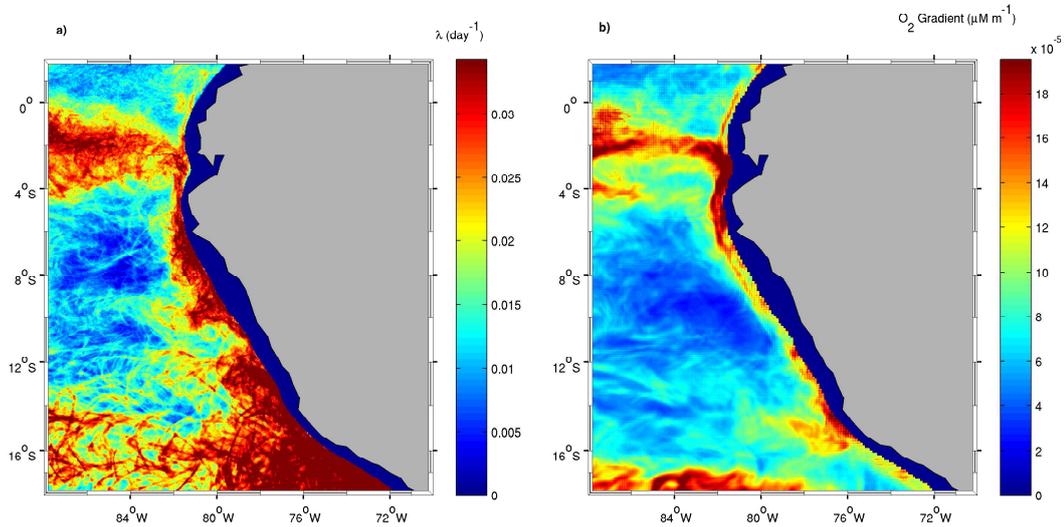

Figure S7: Mean fields averaged over $300-600$ meters depth and during year 21. a) FSLE. b) $O_2$ horizontal gradient.

## Physical and biogeochemical variability of $O_2$

Biogeochemical fluxes undergo a sharp decrease in variability around 300 m of depth, as evidenced from Figure 7 of [14]. That figure clearly indicates that the variability of physical fluxes becomes two orders of magnitude larger than that of biogeochemical fluxes below ∼300 m, since the ratio is ∼1% and lower. Therefore, $O_2$ can be considered as a passive tracer below 300 m. The ratio of variabilities (quantified as root mean square values) of physical and biogeochemical fluxes at $12^o$S is shown in Fig. S8. The dataset used is the same as that described in the Methods section of the main text, and physical and biological fluxes are identified from the terms of physical and biological origin in the balance equation for oxygen concentration.

Again, Fig. S8 highlights that below 300 m the variability in biogeochemical

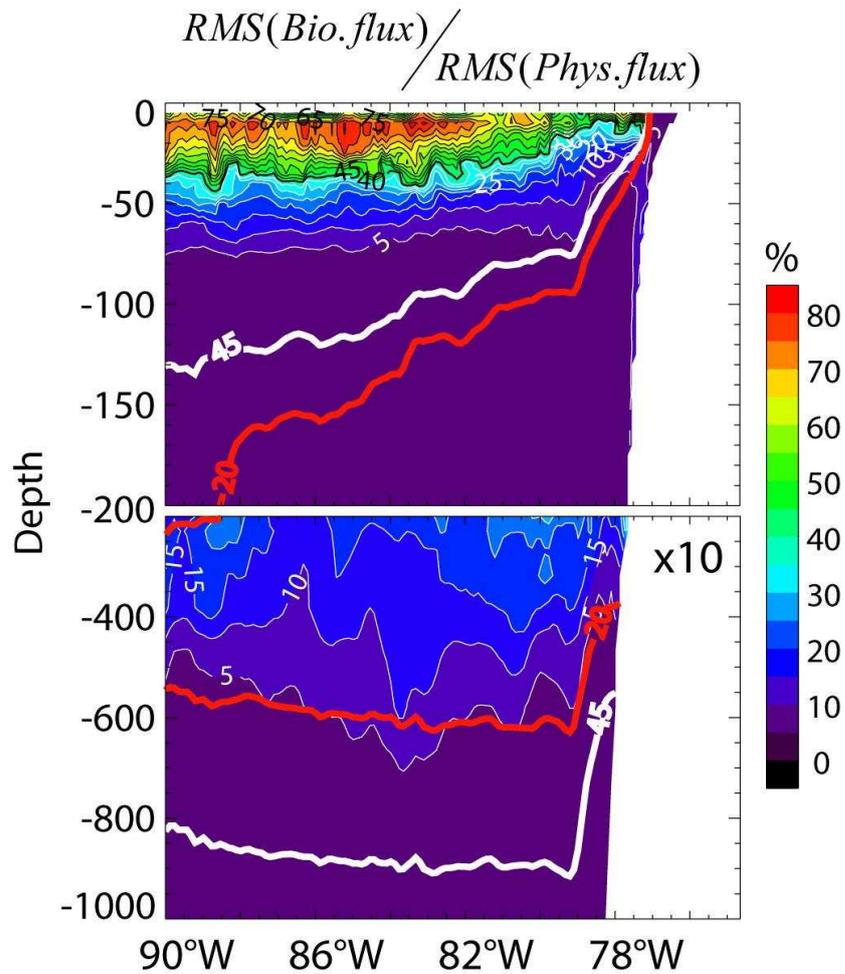

Figure S8: Ratio (in %) of the variability in the biogeochemical fluxes versus the variability of the physical flux along a vertical section at 12°S. The color bar applies to the upper panel, and the colors and values of the ratio in the lower panel correspond to values multiplied by 10 to highlight the vertical variability. The 45 $\mu$M (white line) and 20 $\mu$M (red line) $O_2$ isolines are also shown.

13

fluxes becomes two orders of magnitude smaller than the physical ones. We note that over the whole OMZ, variability in physical fluxes remains in general larger than the variability in biogeochemical fluxes, but small biological changes in O$_2$ can still diminish correlations between physical stirring and O$_2$. Below 300 m the ratio between fluxes drops below 1% so that one can consider that below this depth the physical fluxes strongly dominate over biogeochemical processes. This is consistent with our observations, reported in the main text, of strong correlations between physical stirring and O$_2$ distributions in the range 380-600 m, whereas this correlation is not so strong at upper leyers. To illustrate what is happening near the OMZ boundaries, we calculated the proportion of explained variance by the physical fluxes at different depths. The proportion of total variance ($Var$) in the O$_2$ change rate explained by the physical flux can be written (using that the total flux is the sum of the physical and the biogeochemical ones) as [15]:

$$R_{Phys}^2 = 1 - \frac{\text{Var[Biogeochemical flux]}}{\text{Var[Total flux]}}. \quad (1)$$

$R_{Phys}^2$ is shown in Fig. S8 at two depths (112m and 465m) corresponding to different zones of the OMZ. At 112 m, the biogeochemical fluxes, although having lower amplitude than the physical ones, can still contribute to generate local O$_2$ gradients which can influence advection terms. However at 465 m, the biogeochemical fluxes (within the OMZ) are almost two orders of magnitude lower than the physical flux (see Fig. S8 Fig. 7 of [14]). Fig. S9 shows that physical fluxes dominate over most of the study region, but they are more intense near the OMZ latitudinal boundaries. This is more remarkable at 465m than at 112m, which indicates that the lower correlation between the FSLE field and mean O$_2$ gradient at the surface may arise from the relative larger contribution of biogeochemical fluxes to the rate

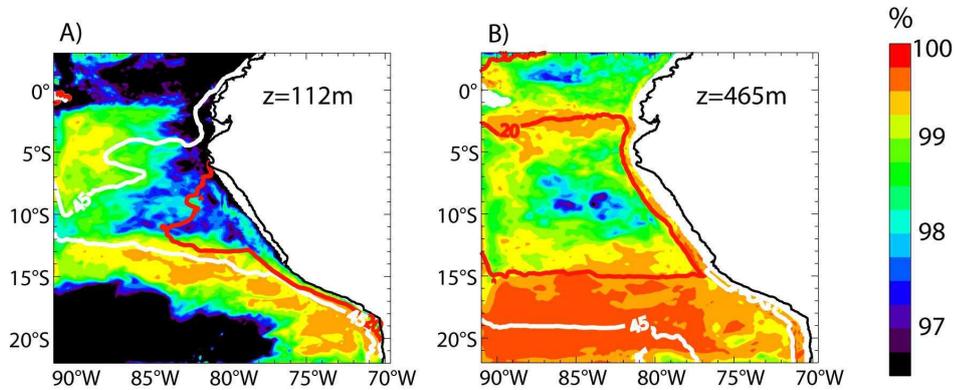

Figure S9: Explained variance of the physical flux with respect to the rate of $O_2$ change in % at A) 112m and B) 465m. The white (red) thick line indicates the iso-oxygene contour at 45 $\mu$M (20 $\mu$M)

of $O_2$ change there.

## Eddy fluxes of $O_2$

We computed $O_2$ eddy fluxes by using the Reynolds transport theorem and the decomposition of the velocity and $O_2$ concentration fields in their mean and fluctuating components. Other physical fluxes such as small-scale turbulent diffusion turn our to be smaller. The eddy flux across each of the mean $20\mu M$ north and south boundaries was computed by averaging the product of the fluctuating velocity component normal to the boundary in the horizontal and in the vertical directions and the fluctuating $O_2$ concentration component, thereby obtaining horizontal and vertical eddy fluxes. The sign of the normal was chosen so that a positive horizontal flux indicates transport towards the interior of the OMZ. For comparison purposes, mean horizontal and vertical fluxes were also computed from the mean velocity normal to the boundaries and the mean $O_2$ concentration. The integration

of the fluxes over the whole frontier gives the flow rate of $O_2$ into the OMZ through those boundaries.

The profiles of mean and eddy horizontal fluxes across the 20 $\mu M$ OMZ boundaries for depths between 200 and 600 meters were computed for simulation year 21 and shown in Fig. S10. The fluxes are averaged also along the horizontal extension of the boundary at each depth. Blue lines are for the northern boundary and red for the southern one. Overall, the mean fluxes are much smaller than the eddy ones. In the northern boundary, eddy fluxes and mean fluxes are both positive up to about 250 m depth highlighting the oxygen supply role the Equatorial undercurrent and the primary southern subsurface counter-current play for the Peru-Chile under-current. At deeper locations the eddy flux is again positive, and the mean flow slightly negative (i.e. pointing outsize the OMZ). At the southern border the eddy flux is positive above 380 m, becoming negligible further down. The mean flux is positive above about 230 m, being negative or negligible below. The total mean horizontal $O_2$ flow rates (Table S1) at the northern and southern boundaries are negative whereas the eddy flow rates are positive. At both OMZ boundaries, mesoscale turbulence tends to transport $O_2$ into the OMZ at a much higher rate than the mean flow removes $O_2$ from the OMZ. At the southern boundary computed flow rates are higher than at the northern boundary probably due to the longer horizontal extension of the southern boundary. The vertical eddy and mean $O_2$ flow rates (Table S2) turn out to be about one thousand times smaller than the horizontal ones, so that the discussion above for the horizontal fluxes applies indeed also to the total fluxes. Vertical fluxes show also magnitude differences between eddy and mean values.

To further investigate the correlation between FSLE and $O_2$ eddy flux, we singled

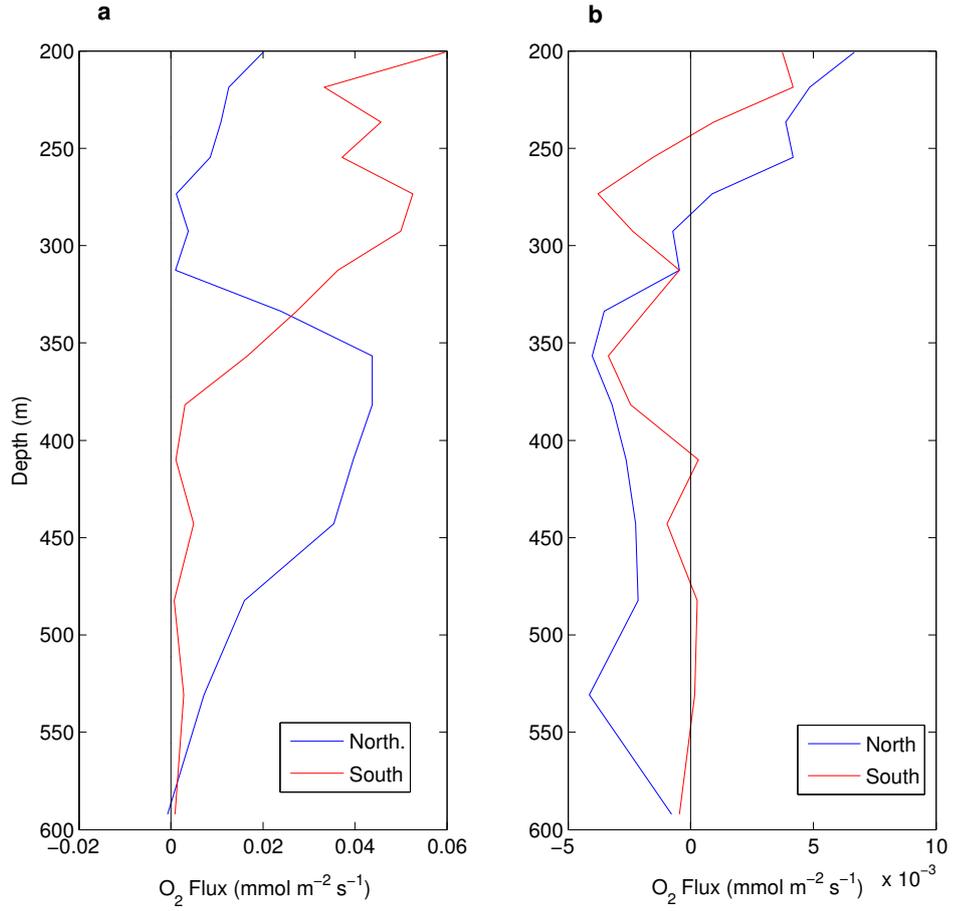

Figure S10: Vertical profiles of horizontal $O_2$ fluxes at the $20\mu M$ OMZ mean boundaries between 200 and 600 m depth. The fluxes have been averaged along the horizontal extension of the boundary (from coast to $88^oW$) for each depth, and during simulation year 21. a) Eddy flux profile (these are the red curves in Fig. 3 of the main text). b) Mean flux profile (note the much smaller values as compared with the eddy fluxes.)

| Boundary | Eddy ($\mu mol\ s^{-1}$) | Mean ($\mu mol\ s^{-1}$) |
|---|---|---|
| Northern | 6.15×10$^6$ | -2.01×10$^5$ |
| Southern | 1.00×10$^7$ | -3.56×10$^5$ |

Table S1: Horizontal eddy and mean O$_2$ flow rates across the 20 $\mu M$ mean boundaries from 200 to 600 m depth and from coast to 88$^o$W, averaged over simulation year 21.

| Boundary | Eddy ($\mu mol\ s^{-1}$) | Mean ($\mu mol\ s^{-1}$) |
|---|---|---|
| Northern | 1197.0 | 16.5 |
| Southern | -4226.4 | -269.7 |

Table S2: Vertical eddy and mean O$_2$ flow rates across the 20 $\mu M$ mean boundaries from 200 to 600 m depth and from coast to 88$^o$W, averaged over simulation year 21.

out the contributions due to intraseasonal variability (timescales between 2-100 days, computed from the departure from monthly mean in the considered year 21) as opposed to the departures from the total annual mean used before. Indeed isolating the sole contribution of the O$_2$ eddy flux due to intraseasonal variability appeared interesting since FSLE might not capture stirring at the long seasonal timescales. In doing so, one finds an improved correlation between FSLE and O$_2$ horizontal eddy flux both at the northern and southern OMZ boundaries (linear correlation coefficient 0.47 versus 0.27 (north) and 0.79 versus 0.77 (south)).

## Episodic ventilation events

We characterize the eventual entrainment of water across the OMZ boundaries by displaying in Fig. 11 Hovmöller plots of the time evolution of the O$_2$ anomaly at the North (panel a) and South (panel b) OMZ 20$\mu M$ boundaries at 410 m depth. The anomaly is the actual O$_2$ concentration minus the mean at that location (20$\mu M$). Since the overall eddy fluxes point towards the OMZ interior, positive anomalies

indicate oxygenated water entering into the OMZ. The slopes in the Hovmöller plots indicate overall westward propagation.

$O_2$ anomalies are more variable for the northern (standard deviation of 4.8 $\mu M$) than for the southern boundary (standard deviation of 3.1 $\mu M$). At the Northern limit the $O_2$ front is more stable meaning that there is an available pool of $O_2$ rich waters readily available to be entrained into the OMZ core by propagating mesoscale structures. At the southern limit the $O_2$ anomaly seems to be dependent on the actual $O_2$ content of travelling eddies that occasionally approach the boundary.

At the northern border the anomalies last longer than at the southern one and remain more coherent during their propagation offshore. In the South, ventilation events are more intermittent. However, during the year under analysis, the greatest episode of $O_2$ anomaly crossing the OMZ at this depth occurred in the southern boundary with peak $O_2$ anomaly of $\sim 17$ $\mu M$ around day 300 and between 600 and 800 km from the coast (this is the episode depicted in Figure 2 of the main text).

## Cross-wavelet spectra

As an alternative methodology to quantify correlations between $O_2$ concentrations and the stirring measure provided by FSLE we conducted a cross-wavelet analysis between the $O_2$ concentration and backward FSLE times series from our simulation dataset at three particular locations in order to identify the dominant time scales of co-variability. The locations were: BNDNORTH_OXYCLINE, located on the northern OMZ boundary (84$^o$W,3$^o$S) and 75 m depth; BNDNORTH_CORE, located in the OMZ core (84$^o$W,3$^o$S) at 410 m depth; and BNDSOUTH_CORE, in the southern boundary (84$^o$W,16$^o$S) at 410 m depth.

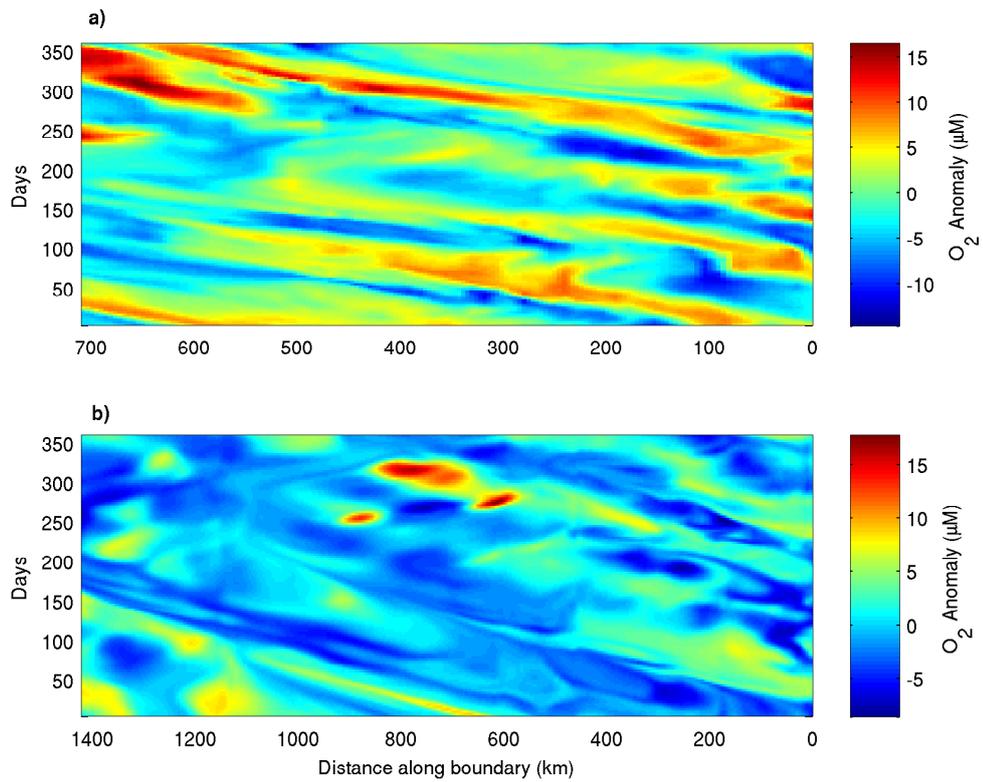

Figure S 11: Hovmöller plots of O$_2$ concentration anomalies at OMZ boundaries at 410 m depth during simulation year 21. a) Northern boundary. b) Southern boundary. Anomaly measured relative to 20 $\mu M$ level. Distance along boundary increases towards offshore, coast is at the right and the anomalies propagate offshore.

The climatological wavelet spectra consist in a wavelet decomposition of the intraseasonal anomaly time series followed by a calculation of the climatology of the wavelet power coefficient at each frequency. Wavelet spectra of $O_2$ and backward FSLE signals were computed following [16] at the three geographical locations. The time span of the signal is 4 years, from 1st July of simulation year 18 to 30 June of simulation year 22. Intraseasonal anomalies were estimated as the departures from the monthly mean. Wavelet power was normalized following Eq. 14 of [16] in order to compare the magnitude of the wavelet power at different frequencies. The mother wavelet used was the Morlet wavelet [16].

The climatology of the resulting cross-wavelet power is shown in Fig. S12. The plots indicate where the energy is dominant, i.e. where is a significant covariance between $O_2$ and FSLE is present, as a function of calendar month. Large energy is found in the intraseasonal frequency band with significant peak energy around 45 days. The marked seasonality of the intraseasonal activity indicates a link with the seasonal modulation of the baroclinic instability of the coastal currents (i.e. the Peru undercurrent) during Austral summer and in the South during Austral winter[17].

Analyzing the dependence of the $O_2$ signal variability with stirring variability, i.e. FSLE levels, is out of the scope of this study. This could in principle be achieved as in [18], namely by artificially decreasing the strength of the non-linear terms in the momentum equation and measuring the resulting change in the variability in $O_2$.

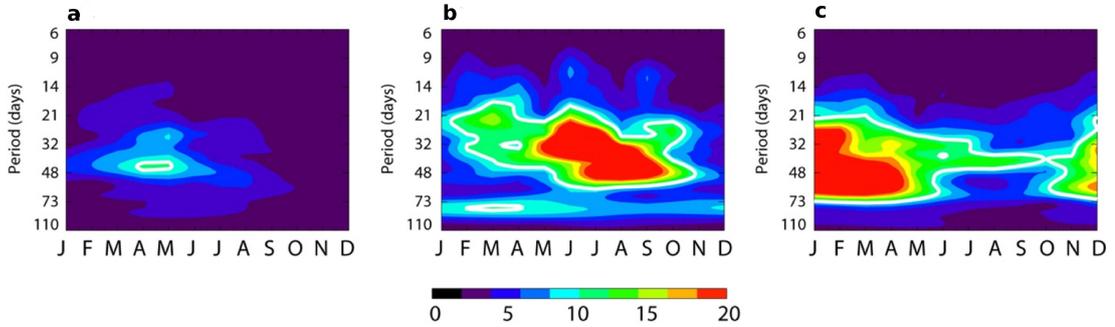

Figure S12: Climatological normalized cross-wavelet spectrum of $O_2$ concentration and FSLE time series at (a) BNDNORTH_CORE, (b) BNDSOUTH_CORE and (c) BNDNORTH_OXYCLINE. The horizontal axis indicates the calendar month. Units of the colorbar are $0.8 \times 10^{-4} M\ day^{-1}$. The white thick contour in all panels indicates the 95% confidence level.



\end{document}